\documentclass[epsig,onecolumn]{mn2e}
\usepackage{epsfig}
\def\beq#1{\begin{equation}\label{#1}}
\def\eeq{\end{equation}}
\def\beqa#1{\begin{eqnarray}\label{#1}}
\def\eeqa{\end{eqnarray}}

\def\epeiso{$E_{\rm p,i}$ -- $E_{\rm iso}$}

\def\sax{{\it Beppo}SAX }
\def\swift{{\it Swift}}
\def\fermi{{\it Fermi}}

\def\h0{H$_{\rm 0}$~}

\title[GRBs Hubble diagram]{The GRBs Hubble diagram in quintessential cosmological models }
\author[M.Demianski et al.]{Marek Demianski$^{1,2,3}$,Ester Piedipalumbo$^{4,5}$, Claudio Rubano$^{4,5}$\\
$^1$ Institute for Theoretical Physics,
University of Warsaw,  Hoza 69, 00-681 Warsaw, Poland  \\
$^2$ Department of Astronomy, Williams College, Williamstown, MA 01267, USA\\
$^3$ Institute for Interdisciplinary Studies "Artes Liberales", Nowy
Swiat 69, 00-046 Warsaw, Poland\\
$^4$ Dipartimento di Scienze Fisiche, Universit\`{a} di Napoli
Federico II, Compl.
Univ. Monte S. Angelo, 80126 Naples, Italy  \\
$^5$ I.N.F.N., Sez. di Napoli, Complesso Universitario di Monte
Sant' Angelo, Edificio G, via Cinthia, 80126 - Napoli, Italy }
\date{Accepted xxx, Received yyy, in original form zzz}
\begin{document}
\maketitle
\begin{abstract}
It has been recently empirically established that some of the
directly observed parameters of GRBs are correlated with their
important intrinsic parameters, like the luminosity or the total
radiated energy. These correlations were derived,  tested
and used to standardize GRBs, i.e., to
derive their luminosity or radiated energy from one or more observables, in order to construct an \textit{ estimated fiducial} Hubble diagram, assuming that radiation propagates in the standard $\Lambda$CDM
cosmological model. We extend these analyses by considering
more general models of dark energy, and an updated data set of high
redshift GRBs. We show that the correlation parameters only weakly
depend on the cosmological model. Moreover we apply a local
regression technique to estimate, in a model independent way, the
distance modulus from the recently updated SNIa sample containing
307 SNIa \cite{SNLS}, in order to calibrate the GRBs 2D correlations,
considering only GRBs with $z \leq 1.4$. The derived calibration
parameters are used to construct a new GRBs Hubble diagram, which we
call the \textit{calibrated} GRBs HD. We also compare the
\textit{estimated} and \textit{calibrated} GRBs HDs. It turns out
that for the common GRBs they are fully statistically consistent,
thus indicating that both of them are not affected by any systematic
bias induced by the different standardizing procedures. We finally
apply our methods to calibrate $95$ long GRBs  with the well-known
Amati relation and construct the \textit{estimated} and
\textit{calibrated}  GRBs Hubble diagram that extends to  redshifts
$z\sim 8$. Even in this case there is consistency between these
datasets. This means that the high redshift GRBs can be used to
test different models of dark energy. We used the
\textit{calibrated} GRBs HD to constrain our quintessential
cosmological model and derived the likelihood values of $\Omega_m$
and $w(0)$.
\end{abstract}
\begin{keywords}
Gamma Rays\,: bursts -- Cosmology\,: distance scale -- Cosmology\,:
cosmological parameters
\end{keywords}
\section{Introduction}
\begin{figure}
\centering{
\includegraphics[width=7cm]{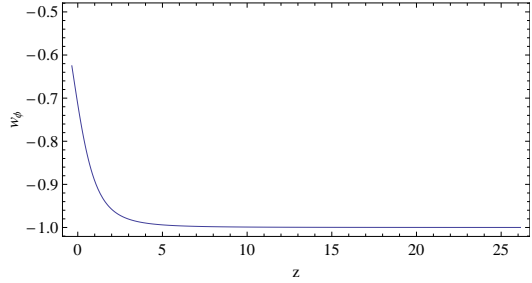}}
\caption{The behaviour of the dark energy equation of state parameter $ w_{\phi}$ on redshift for our fiducial model. It is rapidly changing for $ 0 < z < 5 $ and is almost constant and equal to $ -1 $ at earlier epochs, when the scalar field mimics an effective cosmological constant.}
\label{w}
\end{figure}

Recent observations of high redshift supernovae of type Ia (SNIa) revealed
that the universe is now expanding at an accelerated rate. This
surprising result has been independently confirmed by observations
of small scale temperature anisotropies of the cosmic microwave
background radiation (CMB) \cite{SNLS,Union,SCP,WMAP7}. It is
usually assumed that the observed accelerated expansion is caused by
a so called dark energy, with unusual properties. The pressure of
dark energy $p_{de}$ is negative and it is related with the positive
energy density of dark energy $\epsilon_{de}$ by
$p_{de}=w\epsilon_{de}$, where the proportionality coefficient $w<0$.
According to the present day estimates, about 75\% of matter-energy
in the universe is in the form of dark energy, so that now the dark
energy is the dominating component in the universe. The nature of
dark energy is not known. Proposed so far models of dark energy can
be divided, at least, into three groups: a) a non zero cosmological
constant, in this case $w=-1$, or b) a potential energy of some, not
yet discovered, scalar field, or c) effects connected with non
homogeneous distribution of matter and averaging procedures. In the
last two possibilities, in general, $w$ is not constant and it
depends on the redshift $z$. Observations of type Ia supernovae and
small scale anisotropies of the cosmic microwave background
radiation are consistent with the assumption that the observed
accelerated expansion is due to the non zero cosmological constant.
However, so far the type Ia supernovae have been observed only at
redshifts $z<2$, while in order to test if $w$ is changing with
redshift it is necessary to use more distant objects. New
possibilities opened up when it turned out that some of the Gamma
Ray Bursts are seen at much higher redshifts, the present record is
at $z=8.26$ \cite{Greiner08}. Neither the type Ia supernovae nor the
GRBs are true "standard candles". A type Ia supernova is a final
stage of evolution of a white dwarf, which is a member of a binary
system and accreted enough matter to reach the Chandrasekhar mass
limit of 1.4 $M_{\odot}$. Since the exploding white dwarf always
contains mass of about the Chandrasekhar limit, and white dwarfs
have similar structure, it is reasonable to assume that when they
explode they release approximately  the same amount of energy.
Sterling Colgate already in 1979 suggested that type Ia supernovae
could be used as standard candles \cite{colgate}. When finally, late
in nineteen nineties, astronomers were able to observe very distant
type Ia supernovae, and used them to determine distances to high
redshift galaxies, they discovered that the expansion of the
universe, instead of slowing down as expected, is accelerating.
Because this conclusion has very important consequences it was
prudent to check if  type Ia supernovae have the same intrinsic
luminosity. It turned out that their intrinsic luminosity varies
quite a bit, but with the help of several observational
characteristics it can be estimated with an accuracy of about 15\%.
GRBs are even more enigmatic objects. First of all the mechanism
that is responsible for releasing the incredible amounts of energy
that typical GRB emits is not yet known (see for instance Meszaros
2006 for a recent review on GRBs). It is also not yet definitely
known if the energy is emitted isotropically or is beamed. Despite
of these difficulties GRBs are promising objects that can be used to
study the expansion rate of the universe at high redshifts
\cite{Bradley03,S03,Dai04,Bl03,Firmani05,S07,Li08,Amati08,Ts09}.
Soon after the BeppoSAX satellite was launched in 1996 an afterglow
of now famous GRB970228 has been observed and for the first time it
was possible to determine its redshift and conclusively show that
GRBs occur at cosmological distances. Since then, new satellites
have been launched and it became possible to observationally
determine several important parameters of GRBs. First of all, it is
now possible to determine not only the light curves but also the spectra
of GRB radiation for a wide range of photon energies. Using the
observed spectrum it is possible to derive additional parameters,
for example the peak photon energy $E_{peak}$, at which the burst is
the brightest, and  the variability parameter
$V$ which measures the smoothness of the light curve (for
definitions of these and other parameters mentioned below, see
Schaefer (2007)). From observations of the afterglow it is possible
to derive another set of parameters, redshift is the most important,
and also the jet opening angle $\Theta_{jet}$, derived from the
achromatic break in the light curve of the afterglow, the time lag
$\tau_{lag}$, which measures the time offset between high and low
energy GRB photons arriving at the detector, and $\tau_{RT}$ - the
shortest time over which the light curve increases by half of the
peak flux of the burst. However the most important parameters - the
intrinsic luminosity $L$, and the total radiated energy $E_{\gamma}$
are not directly observable. Assuming that GRBs emit radiation
isotropically it is possible to relate the peak luminosity, $L$, to the observed
bolometric peak flux $P_{bolo}$ by
\begin{equation}
L= 4\pi{d^2}_{L}(z, cp)P_{bolo},
\label{lum}
\end{equation}
 where $d_{L}$ is the
luminosity distance, and $cp$ denotes the set of cosmological
parameters that specify the background cosmological model.
Equivalently one can use the total collimation corrected energy
$E_{\gamma}$ defined by
\begin{equation}
E_{\gamma}=4\pi{d^2}_{L}(z, cp)S_{bolo}F_{beam}(1+z)^{-1},
\end{equation}
where $S_{bolo}$ is the bolometric fluence and
$F_{beam}=(1-\cos{\Theta_{jet}})$ is the beaming factor. It is clear
from these relations that, in order to get the intrinsic luminosity,
it is necessary to specify the fiducial cosmological model and its
basic parameters. But we want to use the observed properties of GRBs
to derive the cosmological parameters. The way out of this
complicated circular situation has been proposed by Schaefer (2007)
and others, who studied correlations between the directly observed
parameters of GRBs ($E_{peak}$, $\tau_{lag}$, $\Theta_{jet}$, and
$V$) and derived parameters like $L$ and $E_{\gamma}$. In his
detailed study of these correlations Schaefer used the standard
$\Lambda$CDM and a quintessence model in which $w$ is redshift
dependent. He showed that the parameters of correlations between $L$
or $E_{\gamma}$ and the observed parameters $E_{peak}$,
$\tau_{lag}$, $\Theta_{jet}$, and $V$ only weakly depend on the
cosmological model. In conclusion his results show that these
correlations can be used to standardize the intrinsic luminosity of
GRB bursts and so to transform them into standard candles.
Recently Basilikos and Perivolaropoulos (2008) have extended the
work of Schaefer, by testing the stability of the correlation
parameters with respect to possible evolution with redshift and
variations of the basic parameters of the background cosmological model. They concluded that
the correlation parameters practically do not depend on the
redshift. They assumed that the background cosmological model is the
standard $\Lambda$CDM model but included the $\Omega_{m}$ parameter
into the set of fitting parameters. In fact they obtained very
similar values of the correlation parameters as those obtained by
Schaefer, but the $\Omega_{m}$ parameter (even at the $1\sigma$ level)
was poorly determined.
Almost at the same time and independently Cardone, Cappoziello and
Dainotti (2009) extended the analysis performed by Schaefer, by
including more objects and adding one recently discovered
correlation between the GRBs X-ray luminosity $L_{X}$ and the time
$T_{a}$, which is characterizing the late afterglow decay. They
then, using a Bayesian procedure, fitted all the correlations, estimating their parameters,
considering as a fiducial cosmological model the $\Lambda$CDM model,
with parameters derived from the WMAP5 data. Finally to check how
their results depend on the parameters of the background cosmological
model they fitted the correlations on a parameter grid
assuming five different values for $\Omega_{m}$ and allowing the
dark energy equation of state parameter $w$ to vary with $z$ as
$w(z)=w_{0}+w_{a}z/(1+z)$, where $(w_{0}, w_{a})$ were  restricted to
vary in the range $-1.3 < w_{0} < -0.3$ and $-1< w_{a} < 1$. They
have found that the distance modulus defined as
\begin{equation}
\mu(z)=25 + 5\log{d_L(z, cp)},
\end{equation}
only very weekly depends on the background cosmological model with
variations not exceeding $1\%$. Using a sophisticated statistical
procedure and the SNIa data they were able to
calibrate the correlation parameters of GRB observables and at the
same time to determine basic parameters that specify the background
cosmological model \cite{CCD}.
The presented so far studies of the correlations between observed
parameters of the GRBs used as the background cosmological model
either the standard $\Lambda$CDM model or a model in which the dark
energy equation of state can change by assuming that either $w(z)$
is a linear function of the redshift or that
$w(z)=w_{0}+w_{a}z/(1+z)$. Let us note that in recent years a host
of very different models of dark energy were proposed, with more
complicated dependence of $w(z)$ and most of them are able to fit at
least the SNIa data.
It is then very important to investigate how strongly, if at all,
the correlation parameters depend on the model of dark energy. In
particular it is interesting to ask what happens if $w(z)$ is rapidly
changing in the present epoch and/or when the luminosity distance is
noticeably different from the $\Lambda$CDM model only at $z>2$.
With this aim in mind in this note we present results of our
analysis of correlations between the observed parameters of
the set of GRBs assembled by Schaefer and later extended by
\cite{CCD} and derived parameters, first by assuming that the
background cosmological model
is one of the quintessence models and secondly by considering an
artificial luminosity distance, which at large redshifts gives much
larger distances in comparison with the standard $\Lambda$CDM model.
To study the first problem we selected one of the quintessence
models that we have studied some time ago \cite{rubano,MECC,pavlov,oscill},  and showed that it is  consistent with the basic
cosmological tests, like type Ia supernovae, the observed CMB
temperature anisotropies power spectrum, and basic parameters of the
large scale structure as determined by the 2dFGRS and Sloan Digital
Sky surveys. In this model the dark energy is represented by a self
interacting scalar field with exponential potential which is
minimally coupled to gravity. The parameters of the
potential have been chosen in such a way that the coupled Einstein -
scalar field dynamical equations describing the flat universe can be
analytically integrated, what is clearly of a great advantage in our
computations.  The energy density of
dark energy depends on time and also the dark energy equation of
state parameter $w_{\phi}$ is time dependent. In Fig. \ref{w} we show how the $w_{\phi}$
parameter is changing with the redshift. As is apparent from this
figure $w_{\phi}$ is rapidly changing for $0 < z < 5$ and is almost
constant and equal to $-1$ at earlier epochs. We have picked this
model precisely because of this dramatic difference between the
behaviour of the dark energy in comparison with previously
considered background cosmological models.
The statistical analysis of the correlation relations between the
observed parameters of GRBs and their important intrinsic parameters
- the total intrinsic luminosity $L$ and total emitted energy
$E_{\gamma}$ performed by Schaefer (2007), Basilikos and
Perivolaropoulos (2008) and Cardone et al. (2008) shows that the
correlation between $E_{peak}$ and $E_{\gamma}$ seems to be the most
robust. The total emitted energy is given by equation (2) and, as it
is apparent from this relation, in order to calculate $E_{\gamma}$,
it is necessary to specify the background cosmological model. We
begin our analysis using the quintessense model with  exponential
potential with parameters fixed by fitting this model to the set of
type Ia supernovae data, the power spectrum of CMB temperature
anisotropies and parameters of the observed large scale structure (
for detailed description see Demianski et al. (2005)). Using the
luminosity distance derived from this model, we fit our correlations, estimate their parameters, and then we update the \textit{estimated} GRBs Hubble diagram. Then, to check the robustness of the discovered correlations between some
of the observed parameters of GRBs and their basic intrinsic
properties, we have tested the stability of values of the correlation
parameters by considering an ad hoc definition of the luminosity
distance that gives much larger distances to objects at $z>2$ than
either of the models mentioned above.  Comparing the likelihood
contours in the plane defined by the values of the slope parameter, a, and the intrinsic scatter, $\sigma_{int}$ \footnote{$\sigma_{int}$ is the amount of intrinsic scatter which we expect around the best fit line, due to the \textit{empirical } nature of the studied correlations, as we will clarify in the section \ref{cal}. }, it turns out that our artificial luminosity distance is changing the maximum likelihood values of the
correlation parameters but the difference is not large. However the
situation changes when we consider, in the space of parameters, the regions of  $1\,\sigma $, $2\,\sigma $ and $3 \,\sigma$ of confidence  for our \textit{crazy} model. It turns out that with respect to the same regions constructed for the quintessential model they overlap only at $ 1 \,\sigma$, but differ consistently at higher levels of confidence.

Moreover we apply a local regression technique to estimate, in a model independent way, the
distance modulus from the recently updated SNIa sample containing
307 SNIa, in order to calibrate the GRBs 2D correlations
considering only GRBs with $z \leq 1.4$. The derived calibration
parameters are used to construct a new  \textit{calibrated} GRBs Hubble diagram.
We also compare the \textit{estimated} and \textit{calibrated} GRBs HDs. We finally
apply our methods to calibrate $95$ long GRBs  with the well-known
Amati relation and construct the  \textit{estimated} and
\textit{calibrated}  GRBs Hubble diagram that extends to  redshifts
$z\sim 8$. Even in this case there is consistency between these
datasets. This means that the high redshift GRBs can be used to
test different models of dark energy. We used the
\textit{calibrated} GRBs HD to constrain our quintessential
cosmological model and to  derive the likelihood values of $\Omega_m$

and $w(0)$.
\section{Standardizing the GRBs and constructing the Hubble diagram}\label{cal}
Following \cite{S07,CCD}, we consider six luminosity relations for
GRBs. First, we consider five that relate GRB peak luminosity,
$L$, or the total burst energy in the gamma rays, $E_\gamma$,
to observables of the light curves and/or spectra:
$\tau_{\mathrm{lag}}$ (time lag), $V$
(variability), $E_{\mathrm{peak}}$ (peak of the $\nu F_\nu$
spectrum), and $\tau_{\mathrm{RT}}$ (minimum rise time)\footnote{It is worth noting that the $E_{\mathrm{peak}}- E_{\gamma}$ correlation listed below is not physically a 2D correlation, because it involves three obervables: the measured fluence, the spectral peak energy and the break time of the optical afterglow light curve needed to measure the jet opening angle. However it can be treated as a 2D correlation from the statistical point of view (to perform the fit). }:
\begin{eqnarray}
  \label{eq:C1}
  \log \left(\frac{L}{1 \; \mathrm{erg} \; \mathrm{s}^{-1}}\right)
  &=& b_1+a_1 \log
  \left[
    \frac{\tau_{\mathrm{lag}}(1+z)^{-1}}{0.1\;\mathrm{s}}
  \right]
  ,
  \\
  \label{eq:C2}
  \log \left(\frac{L}{1 \; \mathrm{erg} \; \mathrm{s}^{-1}}\right)
  &=& b_2+a_2 \log
  \left[
    \frac{V(1+z)}{0.02}
  \right]
  ,
  \\
  \label{eq:C3}
  \log \left(\frac{L}{1 \; \mathrm{erg} \; \mathrm{s}^{-1}}\right)
  &=& b_3+a_3 \log
  \left[
    \frac{E_{\mathrm{peak}} (1+z)}{300\;\mathrm{keV}}
  \right]
  ,
  \\
  \label{eq:C4}
  \log \left(\frac{E_{\gamma}}{1\;\mathrm{erg}}\right)
  &=& b_4+a_4 \log
  \left[
    \frac{E_{\mathrm{peak}} (1+z)}{300\;\mathrm{keV}}
  \right]
  ,
  \\
  \label{eq:C5}
  \log \left(\frac{L}{1 \; \mathrm{erg} \; \mathrm{s}^{-1}}\right)
  &=& b_5+a_5 \log
  \left[
    \frac{\tau_{\mathrm{RT}}(1+z)^{-1}}{0.1\;\mathrm{s}}
  \right]
  ,
\end{eqnarray}
where the $a_i$ and $b_i$ are fitting parameters. To these known
correlations we added a new correlation
\begin{equation}
\log{[L_X(T_a)]}=b_6 + a_6 \log{\left[T_a/(1 + z)\right]},
\end{equation}
 that was recently
empirically found by Dainotti et al. (2009) and later
confirmed by Ghisellini et al. (2008)  and by Yamazaki (2008).
\subsection{Fitting the correlations and estimating their parameters}
\begin{figure}
\includegraphics[width=7 cm]{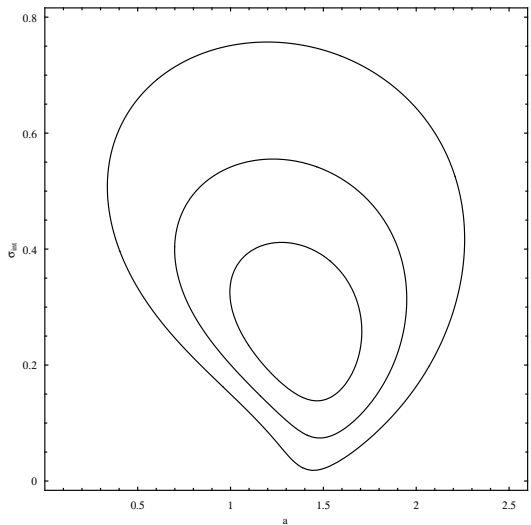}
\caption{Regions of  $68\%$, $95\%$ and $99\%$ of confidence in the
space of parameters $a,\sigma_{int}$.}
\label{confidencereg_egepscal}
\end{figure}

\begin{figure}
\includegraphics[width=7 cm]{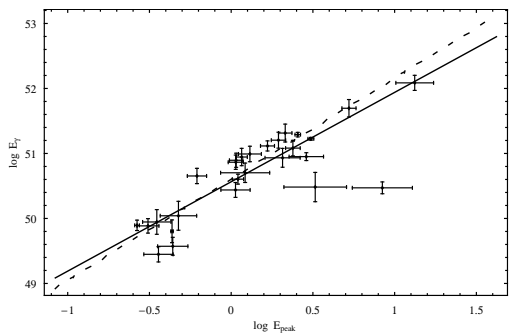}
\caption{Best fit curves for the $E_{peak}-E_{\gamma}$ correlation superimposed on
the data. The solid and dashed lines refer to the results
obtained with the Bayesian and Levemberg\,-\,Marquardt estimators
respectively.}
\label{eg-epcorr_scal}
\end{figure}

\begin{figure}
\includegraphics[width=7 cm]{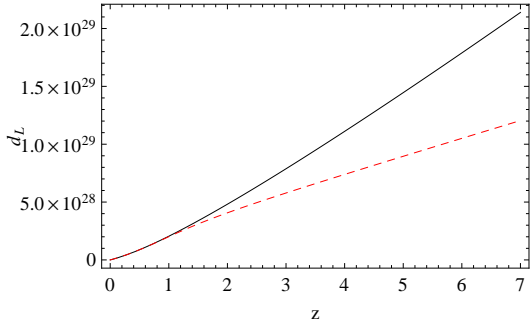}
\caption{The behaviour of the luminosity distance for
our quintessence model (solid line) and the crazy model (dashed
line).}
\label{lumdist}
\end{figure}

\begin{figure}
\includegraphics[width=7 cm]{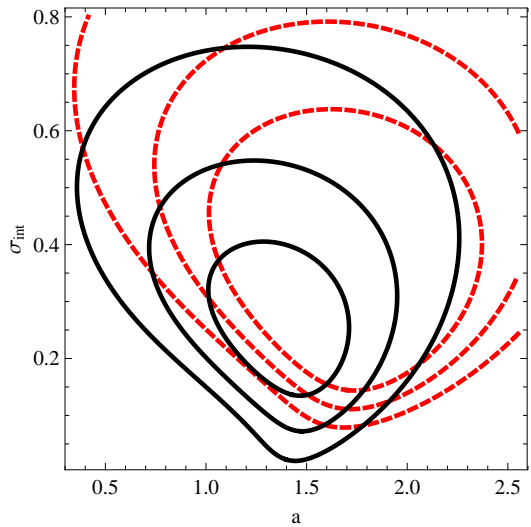}
\caption{Regions of  $68\%$, $95\%$ and $99\%$ of confidence in the
space of parameters $a,\sigma_{int}$ for our \textit{crazy} model (red dashed lines), compared with the same regions of confidence for the quintessential model (solid lines). It turns out that the confidence regions overlap at $1 \sigma$, but   differ consistently at higher levels of confidence. }
\label{confidencereg_egepcrazy}
\end{figure}

\begin{figure}
\includegraphics[width=7 cm]{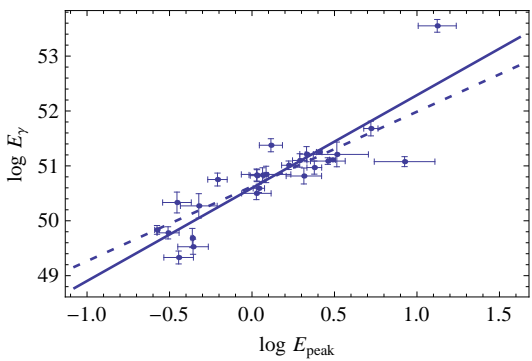}
\caption{Best fit curves for the $E_{peak}-E_{\gamma}$ correlation, for our \textit{crazy} model, superimposed on
the data. The solid and dashed lines refer to the results
obtained with the Bayesian and Levemberg\,-\,Marquardt estimators
respectively.}
\label{eg-epcorr_crazy}
\end{figure}

In fitting these five calibration relations, we need to fit a data
array $\{x_i, y_i\}$ with uncertainties $\{\sigma_{x,i},
\sigma_{y,i}\}$, to a straight line
\begin{equation}
\label{eq: linear} y =b + a x\,.
\end{equation}
Eq.(\ref{eq: linear}) is a linear relation which can be fitted to a
given data set $(x_i, y_i)$ in order to determine the two
fit parameters $(a, b)$\footnote{For the first five correlations we use the
sample of GRBs compiled by Schaefer (2007), while the
$L_X$\,-\,$T_a$ correlation is calibrated using a subsample of the
Willingale et al. (2007) catalog made out of 28 GRBs with measured
redhisft, $\log{L_X(T_a)} \ge 45$ and $1 \le \log{[T_a/(1+z)]} \le
5$ (see \cite{DCC} for the motivation of these limits). Moreover we
note that $F_{beam}$ is calculated with the aid of $d_{L}(z)$: for
small values of $\theta_{ jet}$, the dependence of $F_{beam}$ on
$d_{L}(z)$ is found to be: $F_{beam} \propto d_{L}^{-1/2}$. Using
the data from Schaefer 2007 ($\Omega_{m}=0.27$) we have to multiply
the corresponding $F_{beam}$ value by the following factor:
$[d^{\Lambda}_{L}(0.27,z)/d_{L}(z ,CP )]^{1/2}$}. Actually, the situation is not so
simple since, both the $(y, x)$ variables are affected by
measurement uncertainties $(\sigma_x, \sigma_y)$ which can not be
neglected. Moreover, $\sigma_y/y \sim \sigma_x/x$ so that it is
impossible  to choose as independent variable in the fit the one
with the smallest relative error. Finally, the  correlations we are
fitting are mostly empirical and are not yet derived from an
underlying theoretical model determining the detailed features of
the GRBs explosion and afterglow phenomenology. Indeed, we do expect
a certain amount of intrinsic scatter, $\sigma_{int}$, around the best fit line which
has to be taken into account and determine together with $(a, b)$ by
the fitting procedure. Different statistical recipes are available
to cope with these problems. It is not clear how the fitting technique employed
could affect the final estimate of the distance modulus for a given
GRB from the different correlations so that it is
highly desirable to fit  all of them with the same method.
Following Cardone et al. (2008), we apply a Bayesian motivated
technique \cite{dagostini}  maximizing the likelihood function
${\cal{L}}(a, b, \sigma_{int}) = \exp{[-L(a, b, \sigma_{int})]}$
with\,:
\begin{eqnarray}
L(a, b, \sigma_{int}) & = & \frac{1}{2} \sum{\ln{(\sigma_{int}^2 +
\sigma_{y_i}^2 + a^2
\sigma_{x_i}^2)}} \nonumber \\
~ & + & \frac{1}{2} \sum{\frac{(y_i - a x_i - b)^2}{\sigma_{int}^2 +
\sigma_{x_i}^2 + a^2 \sigma_{x_i}^2}}\,, \label{eq: deflike}
\end{eqnarray}
where the sum is over the ${\cal{N}}$ objects in the sample. Note
that, actually, this maximization is performed in the two parameter
space $(a, \sigma_{int})$ since $b$ may be estimated analytically (solving the equation $\displaystyle \frac{\partial }{\partial b}(L(a, b, \sigma_{int}))=0$, \,as\,:
\begin{equation}
b = \left [ \sum{\frac{y_i - a x_i}{\sigma_{int}^2 + \sigma_{y_i}^2
+ a^2 \sigma_{x_i}^2}} \right ] \left [\sum{\frac{1}{\sigma_{int}^2
+ \sigma_{y_i}^2 + a^2 \sigma_{x_i}^2}} \right ]^{-1}\,. \label{eq:
calca}
\end{equation}
To quantitatively estimate the goodness of this fit we use
the median and root mean square of the best fit residuals, defined
as $\delta = y_{obs} - y_{fit}$ which are computed for the different
correlations we consider.
To quantify the uncertainties of some fit parameter $p_i$, we
evaluate the marginalized likelihood ${\cal{L}}_i(p_i)$  by
integrating over the other parameter. The median value for the
parameter $p_i$ is then found by solving\,:
\begin{equation}
\int_{p_{i,min}}^{p_{i,med}}{{\cal{L}}_i(p_i) dp_i} = \frac{1}{2}
\int_{p_{i,min}}^{p_{i,max}}{{\cal{L}}_i(p_i) dp_i} \ . \label{eq:
defmaxlike}
\end{equation}
The $68\%$ ($95\%$) confidence range $(p_{i,l}, p_{i,h})$ are then
found by solving \cite{dagostini}\,:
\begin{equation}
\int_{p_{i,l}}^{p_{i,med}}{{\cal{L}}_i(p_i) dp_i} = \frac{1 -
\varepsilon}{2} \int_{p_{i,min}}^{p_{i,max}}{{\cal{L}}_i(p_i) dp_i}
\ , \label{eq: defpil}
\end{equation}
\begin{equation}
\int_{p_{i,med}}^{p_{i,h}}{{\cal{L}}_i(p_i) dp_i} = \frac{1 -
\varepsilon}{2} \int_{p_{i,min}}^{p_{i,max}}{{\cal{L}}_i(p_i) dp_i}
\ , \label{eq: defpih}
\end{equation}
with $\varepsilon = 0.68$ and $\varepsilon = 0.95$ for the $68\%$
and $95\%$ confidence level. Just considering  the correlation
relation between $E_{\gamma}$ and $E_{peak}$ of the form
\begin{equation}
 \log{E_{\gamma}}= a\log{E_{peak}} + b\,,
 \label{eq: corr}
\end{equation}
we find that   the likelihood method gives $a=1.37$, $b=50.56$ and
$\sigma_{int}=0.25$. In Fig. \ref{confidencereg_egepscal} we show
the likelihood contours in the $(a, \sigma_{int})$ plane and
in Fig. \ref{eg-epcorr_scal} we show the correlation between the
observed $\log{E_{peak}}$ and derived $\log{E_{\gamma}}$ with our
assumed background cosmological model. The solid line is the best
fit obtained using the D'Agostini's method \cite{dagostini} and the
dashed line is the best fit obtained by the weighted $\chi^2$
method. If one marginalizes with respect to $b$, then the likelihood
values $a$ and $\sigma$ are $a=1.39$ and $\sigma=0.27$. In Table 1
we present data for the other correlation relations. It is apparent
that the correlation parameters only weakly depend on the assumed
background cosmological model. Let us stress that in our model the
dark energy equation of state is changing and the $w(z)$ parameter
is quite different from the models considered by Schaefer (2007),
Basilikos and  Perivolaropoulos (2008), and Cardone et al. (2008).
To check the robustness of the discovered correlations between some
of the observed parameters of GRBs and their basic intrinsic
properties we have tested the stability of values of the correlation
parameters by considering an ad hoc definition of luminosity
distance  that places objects with $z>2$ at much larger distances
than either of the models mentioned above (we call this model a
crazy model). First of all in Fig. \ref{lumdist}  we show the
luminosity distance for our quintessence model (solid line) and the
crazy model (dashed line). In this case using the likelihood
minimalization method we get $a=1.69$, $b=50.60$ and
$\sigma_{int}=0.30$. In Fig. \ref{confidencereg_egepcrazy} we show the $68\%$, $95\%$ and $99\%$ of confidence in the
space of parameters $a,\sigma_{int}$ for our \textit{crazy} model (red dotted lines), compared with the same regions of confidence for the quintessential model: it is apparent that our artificial luminosity distance is changing the values of
the correlation parameters and the regions of confidence. In
Fig. \ref{eg-epcorr_crazy} we show the correlation between the
observed $\log{E_{peak}}$ and derived $\log{E_{\gamma}}$ with our
assumed luminosity distance function. The statistical analysis of
the correlation relations between the observed parameters of GRBs
and their important intrinsic parameters shows that all the
relations have a similar statistical weight since both
$\delta_{med}$  and $\delta_{rms}$ have almost the same values over
the full set, where $\delta =y_{obs} - y_{fit}$. This analysis shows
only a modest preference for the $E_{\gamma}$\,-\,$E_p$,
$L$\,-\,$\tau_{lag}$ and $L_X$\,-\,$T_a$ correlations. Taken at face
values, the maximum likelihood estimates for $\sigma_{int}$ favour
the $E_{\gamma}$\,-\,$E_p$ correlation, which seems the most robust,
but when one takes into account the $68\%$ and $95\%$ confidence
ranges all the correlations overlap quite well. In our analysis we
have assumed that the fit parameters do not change with the
redshift, which indeed spans a quite large range (from $z = 0.125$
up to $z = 6.6$). The limited number of GRBs prevents detailed
exploration of the validity of this usually adopted working
hypothesis, which we tested somewhat investigating if the residuals
correlate with the reshift. We have not found any significant
correlation. Moreover we tested the fit of the
$E_{\gamma}$\,-\,$E_p$ correlation with respect to the evolution with redshift,
separating the GRB samples  into four groups corresponding to the
following redshift  bins: $z\in [0,1]$, $z\in [1,2]$, $z\in [2,3]$
and $z\in [3,7]$. We thus maximized the likelihood in each group of
redshifts and determined the best fit parameters $a$, $b$
with $1\sigma$ errors and the intrinsic dispersion $\sigma_{int}$.
It turns out that no statistical evidence of a dependence of the
$(a, b,\sigma_{int})$ parameters on the redshift exists. This is in
agreement with what has recently been found by Cardone et al. (2008)
and Basilakos \& Perivolaropoulos (2008).
\subsection{Constructing the Hubble diagram}
Once the six  correlations have been fitted, we can now use them
to construct the estimated GRBs Hubble diagram. In order to reduce the error
and, in a sense, marginalize over the possible systematic biases
present in each of the correlations, we take a weighted average of
the distance modulus provided by each of the six 2D laws in Table 1.
Just as an example, let us remind that the luminosity distance
of a GRB with redshift $z$ may be computed as\,:
\begin{equation}
d_L(z) = \left \{
\begin{array}{l}
\displaystyle{E_{\gamma} (1 + z)/4 \pi F_{beam} S_{bolo}}\,, \\ ~ \\
\displaystyle{L/4 \pi P_{bolo}}\,, \\ ~ \\
\displaystyle{L_X (1 + z)^{\beta_a + 2}/4 \pi F_X(T_a)}\,, \\
\end{array}
\right. \ \label{eq: dlval}
\end{equation}
depending on whether some correlation is used.

\begin{figure}
\includegraphics[width=7 cm]{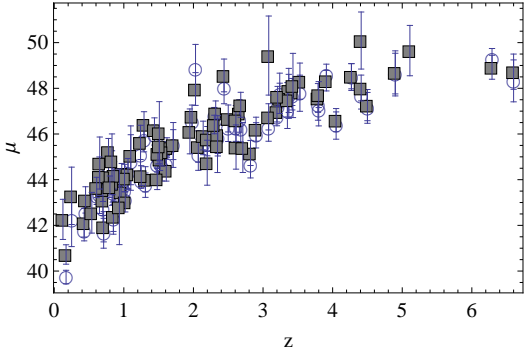}
\caption{The \textit{calibrated} (empty circles) and \textit{estimated} quintessential (full boxes) Hubble diagram.}
\label{hdgrb}
\end{figure}

The uncertainty of $d_L(z)$ is then estimated through the propagation of the measurement errors on the involved
quantities. In particular, remembering that  all the 2D correlation relations  can be written as a linear relation, as in Eq. (\ref{eq: linear}), where $y$ is  the distance dependent quantity, while $x$ is not, the error on the distance
dependent quantity $y$ is estimated as:
\begin{equation}
\sigma(\log{y}) = \sqrt{a^2 \sigma^2(\log{x}) + \sigma_{int}^2}\,,
\label{eq: siglogy}
\end{equation}
and is then added in quadrature to the uncertainties on the other terms entering
Eq.(\ref{eq: dlval}) to get the total uncertainty.  The distance modulus $\mu(z)$ is
easily obtained from its definition\,:
\begin{equation}
\mu(z) = 25 + 5 \log{d_L(z)}\,, \label{eq: defmu}
\end{equation}
with its uncertainty obtained again by error propagation. Following
Schaefer (2007) and Cardone et al. (2008), we finally estimate the
distance modulus for the $i$\,-\,th GRB in the sample at redshift
$z_i$ as\,:
\begin{equation}
\mu(z_i) = \left [ \sum_{j}{\mu_j(z_i)/\sigma_{\mu_j}^2} \right ] \
{\times} \ \left [ \sum_{j}{1/\sigma_{\mu_j}^2} \right ]^{-1}\,,
\label{eq: muendval}
\end{equation}
with the uncertainty given by:
\begin{equation}
\sigma_{\mu} = \left [ \sum_{j}{1/\sigma_{\mu_j}^2} \right ]^{-1}\,,
\label{eq: muenderr}
\end{equation}
where the sum runs over the considered empirical laws. Joining the
Willingale et al. (2007) and Schaefer (2007) samples and considering
that 17 objects are in common, we end up with a catalog of 83 GRBs
which we use to build the Hubble diagram plotted in Fig.
\ref{hdgrb}. We will refer in the following to this data set as the
{\it fiducial} GRBs Hubble diagram (hereafter, HD) since to compute the
distances it relies on the calibration based on the fiducial model.
We  applied such a procedure first for our quintessence
model  and then for our artificial \textit{crazy} model.
Although the above analysis has shown that the choice of the
underlying cosmological model has only a modest impact on the final
estimate of the distance modulus, we compared our \textit{estimated fiducial HD} with a
model independent calibrated HD, carried out using SNIa as distance
indicators. Actually the SNIa Hubble diagram gives the
value of $\mu(z)$ for a subset of the GRBs sample with $z \le 1.4$
which can then be used to calibrate the  correlations (see
\cite{CCD} for details). Assuming again that this calibration is
redshift independent, one can then build up the Hubble diagram at
higher redshifts using the calibrated correlations for the remaining
GRBs in the sample. We implement here a slightly different version
of the local regression approach described in Cardone et al. (2008),
modifying some steps  in order to improve the computing time (in
particular we use a different choice of the weight function and the
order of the fitting polynomial-taking ). Once we estimated the
distance modulus at redshift $z$ in a model independent way, we can
now calibrate the GRBs correlations considering only GRBs with $z
\le 1.4$ in order to cover the same redshift range spanned by the
SNIa data. For such subset of GRBs we apply the Bayesian fitting
procedure described above to construct a new GRBs Hubble diagram
that we call the {\it calibrated} GRBs HD. Plotted in
Fig.\,\ref{hdgrb}, this HD now contains only the 69 Schaefer GRBs
since we have not used the $L_X$\,-\,$T_a$ correlation because it is
impossible to calibrate it with the local regression based method.
It turns out that the \textit{calibrated} and \textit{estimated} data sets,
although containing a different number of objects (83 vs 69), are consistent
(see Figs. \,\ref{hdgrb}  ).
 This qualitative agreement is also confirmed by considering
the numbers of GRBs that are more than $2 \sigma$ away from the
theoretical fiducial HD.
\section{Constructing the Hubble diagram and fitting the $E_{\rm p,i}$ -- $E_{\rm iso}$ correlation}
\begin{figure}
\includegraphics[width=6 cm]{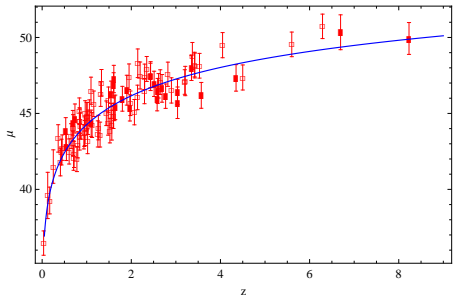}
\caption{The \textit{calibrated} GRBs Hubble diagram with overplotted the
distance modulus predicted by the fiducial  model. The full boxes correspond to the data set
of 70 GRBs compiled by Amati et al. 2008, while the empty ones correspond to the added sample of 20
GRBs compiled by Amati et al. 2009. } \label{hdfidamat}
\end{figure}

\begin{figure}
\includegraphics[width=6 cm]{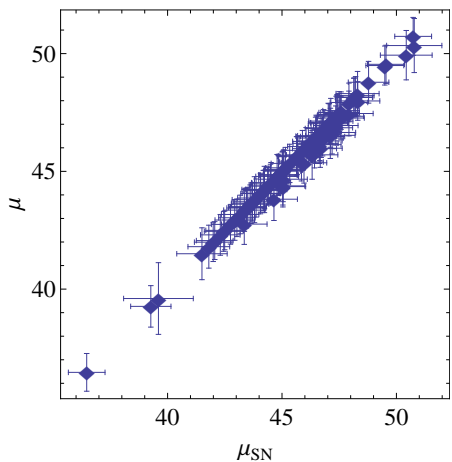}
\caption{Comparison of the distance modulus $\mu(z)$ for the
\textit{calibrated} and \textit{estimated} GRBs Hubble diagram made up fitting the Amati correlation.}
\label{hdcalfidamat}
\end{figure}

\begin{figure}
\includegraphics[width=6cm]{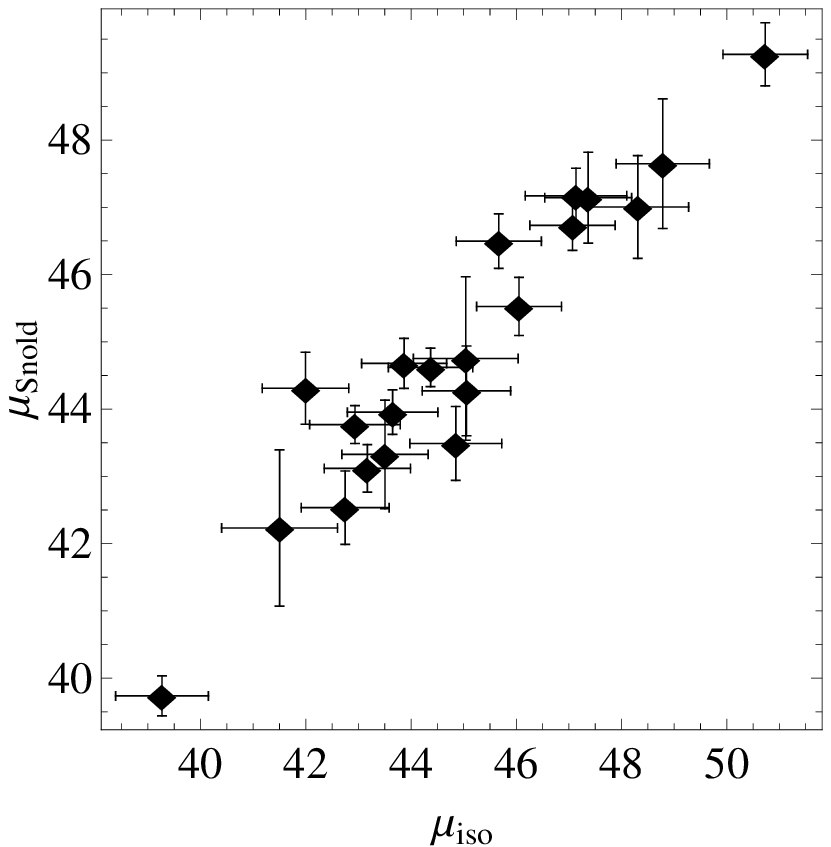}
\caption{Comparison of the distance modulus $\mu(z)$ for the \textit{calibrated}  GRBs
Hubble diagram made up by fitting the five $2-D$ correlations illustrated above, $\mu_{SNold}$,  and the \textit{estimated} GRBs Hubble diagram made up by fitting the Amati correlation, $\mu_{iso}$, limited
to the 21 GRBs that appear in both the Schaefer and Amati samples (Schaefer 2007, Amati et al. 2008, Amati et al. 2009).
It turns out that these datasets are fully consistent. }
\label{hdsch_iso}
\end{figure}

In this section we investigate the possibility of constructing the
Hubble diagram from the $E_{\rm p,i}$ -- $E_{\rm iso}$ correlation,
here  $E_{\rm p,i}$ is the peak photon energy of the intrinsic
spectrum and $E_{\rm iso}$ the isotropic equivalent radiated energy. This correlation was initially discovered on a small
sample of \sax{} GRBs with known redshifts \cite{Amati02} and
confirmed afterwards by Swift observations \cite{Amati06}. Although
it was the first correlation discovered for GRBs  it was never used
for cosmology because of its significant "extrinsic" scatter.
However, the recent increase in the efficiency of GRB  discoveries
combined with the fact that $E_{\rm p,i}$ -- $E_{\rm iso}$
correlation needs only two parameters that are directly inferred
from observations (this fact minimizes the effects of systematics
and increases the number of GRBs that can be used, by a factor $\sim
3$) makes the use of this correlation an interesting tool for
cosmology. Previous analyses of the \epeiso{} plane of GRBs
showed that different classes of GRBs exhibit different behaviours,
and while normal long GRBs and X--Ray Flashes (XRF, i.e.
particularly soft bursts) follow the $E_{\rm p,i}$ -- $E_{\rm iso}$
correlation, short GRBs and the peculiar very close and
sub--energetic GRBs do not (Amati et al. 2008). This fact depends on
the different emission mechanisms involved in the two classes of
events and makes the \epeiso{} relation a useful tool to distinguish
between them \cite{Antonelli09}. A natural explanation for the
short/long dichotomy and the different locations of these classes of
events in the \epeiso{} plane is provided by the "fireshell" model
of GRBs \cite{Ruffini09}. The impact of selection and instrumental
effects on the \epeiso{} correlation of long GRBs was investigated
since 2005, mainly based on the large sample of BATSE GRBs with
unknown redshifts. Different authors came to different conclusions
(see for instance \cite{Ghirlanda05}). In particular,
\cite{Ghirlanda05} showed that BATSE events potentially follow the
\epeiso{} correlation and that the question to clarify is if, and
how much, its measured dispersion is biased. There were also claims
that a significant fraction of \swift{} GRBs is inconsistent with
this correlation \cite{Butler07}. However,  when considering those
\swift{} events with peak energy measured by broad--band instruments
like, e.g., Konus--WIND or the \fermi/GBM  or reported by the BAT
team in their catalog \cite{Sakamoto08} it is found that they are
all consistent with the \epeiso{} correlation as determined with
previous/other instruments \cite{Amati09}. In addition, it turns out
that the slope and normalization of the correlation based on the single data
sets provided by GRB detectors with different sensitivities and energy
bands are very similar. These facts further support
the reliability of the correlation \cite{Amati09}. In our
analysis we considered the sample of 95 long GRB/XRF only, compiled
in (Amati et al. 2008) and (Amati et al. 2009). The redshift
distribution covers a broad range of $z$, from $0.033$ to $8.23$,
thus extending far beyond that of SNIa (z $<$$\sim$1.7), and
including GRB $092304$, the new high-z record holder of Gamma-ray
bursts. As above we apply a Bayesian technique \cite{dagostini} to
fit the  \epeiso{} correlation, maximizing the likelihood
function of Eq. (\ref{eq: deflike}). It turns out that our
likelihood method gives $a=1.56$, $b=52.77$ and $\sigma_{int}=0.31$.
\begin{figure}
\includegraphics[width=7 cm]{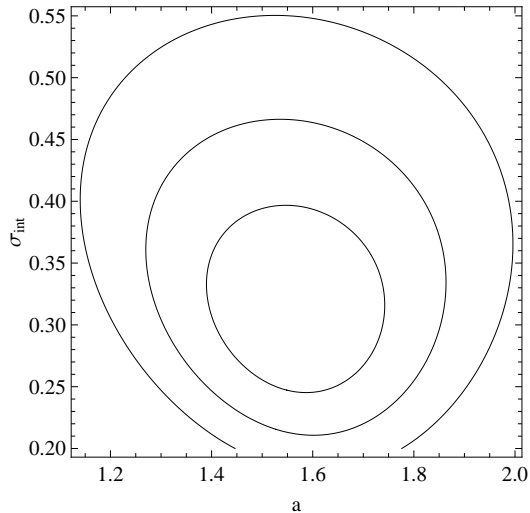}
\caption{Regions of  $68\%$, $95\%$ and $99\%$ of confidence in the
space of parameters $a,\sigma_{int}$, relatively to the fit of the Amati correlation.}\label{confisoamati}
\end{figure}
\begin{figure}
\includegraphics[width=7 cm]{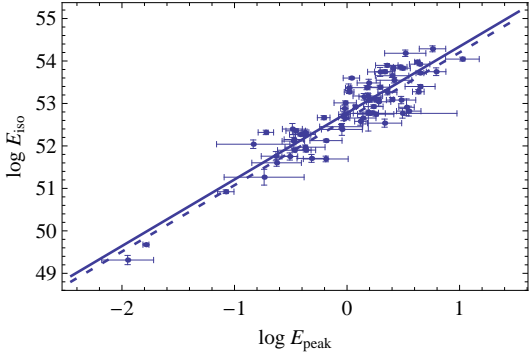}
\caption{Best fit curves, relatively to the fit of the Amati correlation, superimposed on
the data: the solid and dashed lines refer to the results
obtained with the Bayesian and Levemberg\,-\,Marquardt estimators
respectively.}\label{coriso}
\end{figure}
The likelihood contours in the $(a, \sigma_{int})$ plane are shown
in Fig. \ref{confisoamati} and finally in Fig. \ref{coriso} we show
the correlation between the observed $\log{E_{peak}}$ and derived
$\log{E_{iso}}$ with our assumed background cosmological model. The
solid line is the best fit obtained using the D'Agostini's method
\cite{dagostini} and the dashed line is the best fit obtained by the
weighted $\chi^2$ method. Marginalizing over $b$, the likelihood
values for the parameters $a, \sigma$  are $a=1.57$ and
$\sigma=0.32$ respectively.
As in the previous cases we recalibrate also the $E_{\rm p,i}$ --
$E_{\rm iso}$ correlation using the SNeIa data. Once such
correlations have been calibrated, we can now use them to compute the
\textit{estimated} and  {\it calibrated} GRBs Hubble diagram,
according to the relations:
\begin{eqnarray}
\label{amatihd}
d^2_L(z)&=&{E_{iso} (1 + z)\over 4 \pi  S_{bolo}}\,,\\
\mu(z) &=& 25 + 5 \log{d_L(z)}.\nonumber
\end{eqnarray}
It turns out that there are seven GRBs deviating from the fiducial
distance modulus $\mu$ more than $2\sigma$; moreover also for the
Amati correlation the \textit{estimated} and \textit{calibrated} HDs are consistent, as
shown in  Figs.\,\ref{hdfidamat} and
\ref{hdcalfidamat}. Moreover, in order to test the reliability
of the Amati correlation with respect to the other correlations, we compare the distance modulus $\mu(z)$ for the
\textit{calibrated} Hubble diagram constructed of GRBs dataset
used for fitting the five $2D$ correlations illustrated
above and the \textit{calibrated} Hubble diagram constructed of
GRBs dataset used for fitting the Amati correlation. However there are only  21
GRBs that appear in both the Schaefer \cite{S07} and Amati \cite{Amati09} samples.
It turns out that there is full consistency between
these datasets (see Fig. \ref{hdsch_iso}), and the resulting
distances are strongly correlated with the Spearman's correlation
$\rho = 0.92$.
 We observe that our sample includes 4 GRBs with $z > 5$, namely 060522
at $z = 5.11$, 050904 at $z = 6.29$, 060116 at $z = 6.60$ and
$090423$ at $z=8.23$. As is apparent from Fig.\,\ref{hdfidamat},
there is a clear gap between the  $z > 6$ GRBs and the rest of the
sample so that one could wonder whether these  objects share the
same properties with the other GRBs.
Actually, even if extreme in redshift, all of them follow quite well
the considered here correlation. In particular the observed
properties of 090423 are in a good agreement with the Amati
correlation. Therefore, we do not expect that they introduce a bias
in the fiducial HD. Moreover, since they are at such a high
redshift, they are not used in the calibration based on the local
regression analysis. Nevertheless, their distance moduli in the
fiducial and calibrated HDs are consistent. We therefore
conclude that, notwithstanding the doubts on the validity of this 2D
correlation at very high $z$, the inclusion of these three GRBs does
not bias the results. At the end we use the  {\it calibrated} GRBs
Hubble diagram to test if our fiducial cosmological model is able to
describe the background expansion up to  redshifts $z\sim 8$.  In
the Bayesian approach to model testing, we explore the parameter
space through the likelihood function\,:
\begin{equation}
{\cal{L}}_{GRB}({\bf p}) \propto \exp{[-\chi_{GRB}^2({\bf p})/2]}\,,
\label{eq:likegrbsimple}
\end{equation}
with
\begin{equation}
\chi_{GRB}^2({\bf p}) =
\sum_{i = 1}^{{\cal{N}}_{GRB}}{\left [
\frac{\mu_{obs}(z_i) - \mu_{th}(z_i)}{\sqrt{\sigma_i^2 + \sigma_{GRB}^2}} \right ]^2}\,,
\label{eq: defchigrb}
\end{equation}
where $\sigma_{GRB}$ takes into account  the intrinsic scatter inherited from the
scatter of GRBs around the  $E_{\rm p,i}$ -- $E_{\rm iso}$ correlation,
and  {\bf p} denotes the set of model parameters - $H_0,h$ in our case.
The likelihood values are $H_0=0.95\pm 0.03\,\, h=0.62^{+0.07}_{ -0.05}$,
which correspond to $\Omega_m= 0.26^{+0.03}_{-0.04}$ and
$w(0)=-0.77^{+0.03}_{-0.05}$. If we marginalize over $h$ we obtain
$H_0=0.93^{+0.09}_{-0.03}$, which  corresponds to
$\Omega_m= 0.31^{+0.04}_{-0.09}$ and $w(0)=-0.78^{+0.04}_{-0.09}$.

\section{Discussion and Conclusions}
 Recently several interesting correlations among the Gamma Ray Burst
(GRB) observables  have been identified. Proper evaluation and
calibration of these correlations are needed to use the GRBs as
standard candles constraining the expansion history of the universe
up to redshifts of $z\sim 8$. Here we used the  GRB data set
recently compiled by Schaefer (2007) to investigate, in a
quintessential cosmological scenario, the fitting of six 2D
correlations: namely $\tau_{\mathrm{lag}}$ (time lag), $V$
(variability), $E_{\mathrm{peak}}$ (peak of the $\nu F_\nu$
spectrum), and $\tau_{\mathrm{RT}}$ (minimum rise time). In
particular we have investigated the dependence of the calibration,
on the assumed cosmological model and  a possible evolution of the
correlation parameters  with the redshift. To check the robustness
of the discovered correlations between some of the observed
parameters of GRBs and their basic intrinsic properties, we have
first tested the stability of values of the correlation parameters
by considering an ad hoc definition of luminosity distance that
gives much larger distances to objects at $z>2$ than either of the
fiducial models. It is apparent that our artificial luminosity
distance is changing the values of the correlation parameters,
however the difference is not dramatic at $1 \sigma$ of confidence,
but becomes significant at higher levels. We then investigated the
effects of cosmological parameters on the calibration parameters
$(a, b, \sigma_{int})$ when we explore the parameters-space of our
fiducial cosmological model. It turns out that they very weakly
depend on the assumed values of the cosmological parameters.
Moreover our analysis shows only a modest preference for the
$E_{\gamma}$\,-\,$E_p$, $L$\,-\,$\tau_{lag}$ and $L_X$\,-\,$T_a$
correlations. Taken at face values, the maximum likelihood estimates
for $\sigma_{int}$ favour the $E_{\gamma}$\,-\,$E_p$ correlation,
which seems the most robust, but when one takes into account the
$68\%$ and $95\%$ confidence ranges all the correlations overlap
quite well. In our analysis we have assumed that the correlation
parameters do not change with the redshift, which indeed spans a
quite large range (from $z = 0.125$ up to $z = 6.6$). The limited
number of GRBs prevents detailed exploration of the validity of this
usually adopted working hypothesis, which we tested somewhat by
investigating if the residuals correlate with the reshift.  Moreover
we tested the correlation $E_{\gamma}$\,-\,$E_p$ with respect to the
evolution with redshift, separating the GRB samples  into four
groups corresponding to the following redshift  bins: $z\in [0,1]$,
$z\in [1,2]$, $z\in [2,3]$ and $z\in [3,7]$. We thus maximized the
likelihood in each bin of redshift and determined the best fit
calibration parameters $a$, $b$ with $1\sigma$ errors and the
intrinsic dispersion $\sigma_{int}$. We have not found any
statistical evidence of a dependence of the $(a, b,\sigma_{int})$
parameters on the redshift, as recently also reported by Cardone et
al. (2008) and Basilakos \& Perivolaropoulos (2008). Once the six
correlations have been fitted, we constructed the GRBs Hubble
diagram, taking a weighted average of the distance modulus provided
by each of the six 2D correlations considered. Although the above
analysis has shown that the choice of the underlying cosmological
model has only a modest impact on the final estimate of the distance
modulus,  we apply a local regression technique to estimate,
in a model independent way, the distance modulus  from the most
updated SNIa sample, in order to calibrate the GRBs 2D correlations, including only GRBs with $z \leq 1.4$. The derived calibration
parameters have been used to construct a new GRBs Hubble diagram,
which we call the \textit{calibrated} GRBs HD. We finally compare
the \textit{estimated} and \textit{calibrated} GRBs HDs. It turns out
that  for the common GRBs they are fully statistically consistent
thus indicating that both of them are not affected by any systematic
bias induced by the different fitting procedures.  This means
that the high redshift GRBs can be used to test different models of
dark energy.  We also constructed the Hubble diagram using the
$E_{\rm p,i}$ -- $E_{\rm iso}$ correlation, which, recently has been shown to be useful in cosmological considerations.  In our analysis we
used the sample of 95 long GRB/XRF only, compiled in (Amati et
al. 2008) and (Amati et al. 2009), which covers a broad range of
$z$, and includes GRB $090423$, the new high-z record holder of
Gamma-ray bursts. As above we calibrated this correlation
maximizing our likelihood function. It turns out that our
likelihood method gives $a=1.56$, $b=52.77$ and $\sigma_{int}=0.31$.
Applying the same Bayesian technique we recalibrate also  the
$E_{\rm p,i}$ -- $E_{\rm iso}$ correlation using the SNIa data. It
turns out that there are seven GRBs deviating from the fiducial
distance modulus $\mu$ more than $2\sigma$; so that also for the
Amati correlation the fiducial and calibrated HDs are consistent. We
finally use the  {\it calibrated} GRBs Hubble diagram to test if our
fiducial quintessential cosmological model is able to describe the
background expansion up to  redshifts $z\sim 8$. The likelihood
values correspond to $\Omega_m= 0.26^{+0.03}_{-0.04}$ and
$w(0)=-0.74^{+0.01}_{0.03}$.
In conclusion one can say that the correlation parameters that
relate some of the observed parameters of GRBs with their basic
intrinsic parameters only weakly depend on the background
cosmological model. This in turn means that the Gamma Ray Bursts
could be used to derive values of the basic cosmological parameters
and in the future with much larger data set it should be possible to
get more information about the nature of dark energy. Since the
discovered correlations only weakly depend on the background
cosmological model they should be created during the burst itself
and therefore they also provide new constrains on models of the
Gamma Ray Burst.
\begin{table}
\caption{Calibration parameters $(a, b)$, intrinsic scatter $\sigma_{int}$, median, $\delta_{med}$, and root mean square of the best fit residuals, $\delta_{rms}$ for the 2D correlations $log{R} = a \log{Q} + b$ evaluated in two fiducial cosmological models
(the $\Lambda$CDM and the quintessential exponential scalar  field  models) and in our crazy model.
Columns are as follows\,: 1. id of the correlation; 2. number
of GRBs used; 3. maximum likelihood parameters; 4,5 median value and $68$\% and $95\%$ confidence
ranges for the parameters $(a, \sigma_{int})$;  6,7 median and root mean square of the residuals. }
\begin{center}
\begin{tabular}{|c|c|c|c|c|c|c|}
\hline Id & ${\cal{N}}$ & $(a, b, \sigma_{int})_{ML}$ &
$a_{-1\sigma }^{+1\sigma }$ & $(\sigma_{int})_{-1\sigma
}^{+1\sigma}$ &$\delta_{med}$ &
$\delta_{rms}$ \\ \hline \hline
~ & ~ & ~ & ~ & ~&  ~ & \\
$E_{\gamma}^{\Lambda CDM}$\,-\,$E_p$ & 27 & (1.38, 50.56, 0.25) &
$1.37_{-0.26 }^{+0.23 }$ & $0.30_{-0.09 }^{+0.11}$ & $0.01 $&$0.38$\\
~ & ~ & ~ & ~ & ~&  ~ & \\
$E_{\gamma}^{scalar}$\,-\,$E_p$ & 27 & (1.37, 50.56, 0.25) &
$1.37_{-0.26 }^{+0.23 }$ & $0.30_{-0.09 }^{+0.16}$ &$0.01$&$0.38$\\
~ & ~ & ~ & ~ & ~&  ~ & \\
$E_{\gamma}^{crazy}$\,-\,$E_p$ & 27 & (1.36, 50.63, 0.25) &
$1.69_{-0.26 }^{+0.263 }$ & $0.35_{-0.09 }^{+0.12}$ &$0.01$&$0.42$\\
~ & ~ & ~ & ~ & ~&  ~ & \\
$L^{\Lambda CDM}$\,-\,$E_p$ & 64 & (1.24, 52.16, 0.45) &
$1.24_{-0.18 }^{+0.18 }$ & $0.48_{-0.07 }^{+0.07 }$ &$-0.05$&$0.
51$\\
~ & ~ & ~ & ~ & ~&  ~ & \\
$L^{scalar}$\,-\,$E_p$ & 64 & (1.40, 51.95, 0.46) &
$1.24_{-0.18 }^{+0.18 }$ & $0.48_{-0.07 }^{+0.07 }$ &$-0.05$&$0.51$\\
~ & ~ & ~ & ~ & ~&  ~ & \\
$L^{crazy}$\,-\,$E_p$ & 64 & (1.57, 52.31, 0.55) &
$1.57_{-0.21 }^{+0.21 }$ & $0.58_{-0.08 }^{+0.09 }$&$-0.07$&$0.52$ \\
~ & ~ & ~ & ~ & ~&  ~ & \\
$L^{\Lambda CDM}$\,-\,$\tau_{lag}$ & 38 & (-0.80, 52.28, 0.37) &
$-0.80_{-0.14 }^{+0.14 }$ & $0.40_{-0.07 }^{+0.09 }$&$0.02$&$0.40$ \\
~ & ~ & ~ & ~ & ~&  ~ & \\
$L^{scalar}$\,-\,$\tau_{lag}$ & 38 & (-1.13, 52.16, 0.32) &
$-0.80_{-0.14 }^{+0.14 }$ & $0.40_{-0.07 }^{+0.09 }$&$0.02$&$0.4$ \\
~ & ~ & ~ & ~ & ~&  ~ & \\
$L^{crazy}$\,-\,$\tau_{lag}$ & 38 & (-0.90, 52.67, 0.50) &
$-0.90_{-0.18 }^{+0.18 }$ & $0.54_{-0.09 }^{+0.11 }$&$-0.06$&$0.68$ \\
~ & ~ & ~ & ~ & ~&  ~ & \\
$L^{\Lambda CDM}$\,-\,$\tau_{RT}$ & 62 & (-0.89, 52.48, 0.44) &
$-0.89_{-0.18 }^{+0.16 }$ & $0.46_{-0.06 }^{+0.07 }$&$0.04$&$0.47$ \\
~ & ~ & ~ & ~ & ~&  ~ & \\
$L^{scalar}$\,-\,$\tau_{RT}$ & 62 & (-0.89, 52.45, 0.44) &
$-0.89_{-0.16 }^{+0.16 }$ & $0.46_{-0.06 }^{+0.075 }$&$0.03$&$0.48$ \\
~ & ~ & ~ & ~ & ~&  ~ & \\
$L^{crazy}$\,-\,$\tau_{RT}$ & 62 & (-1.07, 52.71, 0.63) &
$-1.079_{-0.21 }^{+ 0.21 }$ & $0.66_{-0.08 }^{+ 0.1 }$&$-0.12$&$0.69$ \\
~ & ~ & ~ & ~ & ~&  ~ & \\
$L^{\Lambda CDM}$\,-\,$V$ & 51 & (1.03, 52.49, 0.48) & $1.04_{-0.29}^{+0.39 }$ & $0.51_{-0.07 }^{+0.09 }$&$-0.09$ &$0.50$\\
~ & ~ & ~ & ~ & ~&  ~ & \\
$L^{scalar}$\,-\,$V$ & 51 & (1.04, 52.47, 0.48) & $1.05_{-0.28}^{+0.29 }$ & $0.51_{-0.07 }^{+0.09 }$&$-0.09$&$0.5$ \\
~ & ~ & ~ & ~ & ~&  ~ & \\
$L^{crazy}$\,-\,$V$ & 51 & (1.42, 52.64, 0.59) & $1.41_{-0.35}^{+0.33 }$ & $0.63_{-0.1 }^{+0.12 }$&$-0.12$&$0.68$ \\
~ & ~ & ~ & ~ & ~&  ~ & \\
$L_X$\,-\,$T_a$ & 28 & (-0.58, 48.09, 0.33) &
$-0.58_{-0.18}^{+0.18 }$ & $0.39_{-0.11 }^{+0.14}$&$-0.12$&$0.43$ \\
~ & ~ & ~ & ~ & ~&  ~ & \\
\hline
\end{tabular}
\end{center}
\end{table}

\subsection*{Acknowledgments}
This paper was supported in part by the Polish Ministry of Science and Higher Education grant NN202-091839.

\end{document}